\documentclass[10pt]{iopart}
\usepackage{iopams}
\usepackage{graphicx,epsfig}
\usepackage{dcolumn}
\usepackage{bm}

 \def\gtap{\mathrel{ \rlap{\raise 0.511ex \hbox{$>$}}{\lower 0.511ex
   \hbox{$\sim$}}}} 
\def\ltap{\mathrel{ \rlap{\raise 0.511ex
    \hbox{$<$}}{\lower 0.511ex \hbox{$\sim$}}}}

\newcommand{\apj}{ApJ}
\newcommand{\apjl}{ApJ Lett.}

\newcommand{\mnras}{MNRAS}

\newcommand{\physrep}{Physics Reports}
\newcommand{\prd}{Phys. Rev. D}

\newcommand{\nat}{Nature}

\begin{document}
\title[Non-Gaussian halo assembly bias]{Non-Gaussian halo assembly bias}
\author{Beth A. Reid$^{1}$, Licia Verde $^{1,2}$, Klaus Dolag$^{3}$, Sabino Matarrese$^{4,5}$, Lauro Moscardini$^{6, 7}$}
$^{1}$ Institute for Sciences of the Cosmos (ICC), University of Barcelona \& IEEC, Barcelona 08028, Spain \\
$^{2}$ ICREA (Instituci\'o Catalana de Recerca i Estudis Avan\c{c}ats) \\   
$^{3}$ Max Planck Institute for Astrophysics, P.O. Box 1317, DÐ85741 Garching, Germany \\
$^{4}$ Dipartimento di Fisica ``G. Galilei'', Universit\`{a} degli Studi di Padova, via Marzolo 8, I-35131 Padova, Italy \\
$^{5}$ INFN, Sezione di Padova, via Marzolo 8, I-35131 Padova, Italy \\
$^{6}$ Dipartimento di Astronomia, Universit\`{a} di Bologna, Via Ranzani 1, I-40127 Bologna, Italy \\
$^{7}$ INFN, Sezione di Bologna, Viale Berti Pichat 6/2, I-40127 Bologna, Italy

\begin{abstract}
The strong dependence of the large-scale dark matter halo bias on the (local) non-Gaussianity parameter, $f_\mathrm{NL}$, offers a promising avenue towards constraining primordial non-Gaussianity with large-scale structure surveys.  In this paper, we present the first detection of the dependence of the non-Gaussian halo bias on halo formation history using  $N$-body simulations.  We also present an analytic derivation of the expected signal based on the extended Press-Schechter  formalism. In excellent agreement with our  analytic prediction, we find that the halo formation history-dependent contribution to the non-Gaussian halo bias (which we call non-Gaussian halo assembly bias) can be factorized in a form approximately independent of redshift and halo mass. The correction to the non-Gaussian halo bias due to the halo formation history can be as large as 100\%, with a suppression of the signal for recently formed halos and enhancement for old halos. This could in principle be a problem for realistic galaxy surveys if observational selection effects were to pick galaxies occupying only recently formed halos.   Current semi-analytic galaxy formation models, for example, imply an enhancement in the expected signal of $\sim 23\%$ and $\sim 48\%$ for galaxies at $z=1$ selected by stellar mass and star formation rate, respectively. 
\end{abstract}

\section{Introduction}
Placing constraints on  deviations from Gaussian primordial fluctuations offers the possibility to test  inflationary models \cite{BKMR04,KomatsuWhitepaper} and probes aspects of inflation (namely the interactions of the inflaton) that are difficult to probe by other means. 

 In this paper we focus on the so-called local non-Gaussianity, which  describes inflation-motivated departures from Gaussian initial conditions  and is   parameterized by  \cite{salopek/bond:1990,  gangui/etal:1994, VWHK00, komatsu/spergel:2001}:
\begin{equation}
\label{eq:fnlnongauss}
\Phi = \phi + f_\mathrm{NL}  (\phi^2 - \left<\phi^2\right>)\,.
\end{equation} 

Here $\phi$ denotes a Gaussian field and  $\Phi$ denotes 
Bardeen's gauge-invariant potential, which on sub-Hubble scales reduces to the usual Newtonian peculiar gravitational potential, up to a minus sign.  The parameter  $f_\mathrm{NL}$ is the amplitude of the non-Gaussian correction; since $\phi \sim 10^{-5}$ and current observational limits restrict $|f_\mathrm{NL}| < 100$ \cite{slosar/etal:2008,komatsu/etal:2010}, we are considering corrections of order $10^{-3}$. 

Recently, Refs.~\cite{dalal/etal:2008b, matarrese/verde:2008} showed that primordial non-Gaussianity affects the clustering
 of dark matter halos (i.e., density extrema), inducing a scale-dependent bias for halos on large scales.
The strong scale-dependence of halo bias ($\propto 1/k^2$) predicted for non-Gaussianity of the local type \cite{dalal/etal:2008b} can provide constraints on $f_\mathrm{NL}$ competitive with those available from  the Cosmic Microwave Background  \cite{dalal/etal:2008b,carbone/verde/matarrese:2008, verde/matarrese:2009}.  Analytic estimates of the amplitude of the scale-dependent bias show good agreement with results from $N$-body simulations \cite{dalal/etal:2008b,grossi/etal:2009, pillepich/porciani/hahn:2010, desjacques/etal:2009}.  

Slosar et al.~(2008) \cite{slosar/etal:2008} argue that the amplitude of the non-Gaussian halo bias should depend on the halo merger history.  Motivated by the idea that quasar activity is triggered by recent mergers, they estimate the amplitude of the non-Gaussian halo bias for recent mergers.  In this paper we extend their reasoning to a more general dependence on the halo merger history through the halo formation redshift $z_f$.  We compare this analysis with the dependence of the non-Gaussian halo bias on halo merger history detected in the $N$-body simulations of Grossi et al.~(2009) \cite{grossi/etal:2009}.  By comparison with the halo merger history dependence of the halo occupation distribution of certain galaxies in semi-analytic models of galaxy formation, we estimate the possible impact of these results on predictions for the amplitude of the non-Gaussian bias in upcoming large scale structure surveys.

This paper is organized as follows. In Section 2 we first revisit the extended Press Schechter non-Gaussian halo merger bias derivation of Ref.~\cite{slosar/etal:2008} and then generalize it to arbitrary halo formation redshifts.  In Section 3 we detect the effect in $N$-body simulations and show the agreement with the analytic description, including the simple halo mass and redshift dependence predicted by the model.  We explore the consequences for  practical determination of $f_\mathrm{NL}$ in Section 4 and we conclude in Section 5. The appendix presents our methodology for fitting our simulation results for the amplitude of the non-Gaussian halo bias mode by mode, i.e. without computing a binned power spectrum.

\section{Theory}
\label{sectheory}
It has been shown \cite{babich/creminelli/zaldarriaga:2004} that  for non-Gaussianity of the local type considered here,  the bispectrum is dominated by the so-called {\it squeezed} configurations, triangles where one wavevector length is much smaller than the other two. In other words, local non-Gaussianity introduces strong coupling between large and small scales.  It is this coupling that alters  halo clustering on large scales (and the halo mass function). In the peak-background split  framework, for Gaussian initial conditions, the short-wavelength modes of the density field are responsible for halo collapse and virialization, while the long wavelength ones modulate halo counts.  The Lagrange bias of halos of mass $M$ at redshift $z_o$ relates their number density (in Lagrange coordinates) to the long wavelength matter overdensity field $\delta_l({\bf x})$ at redshift $z_o$:
\begin{equation}
n_h(M, z_o, {\bf x}) = \bar{n}(M,z_o) (1+b_L(M, z_o) \delta_l({\bf x}, z_o)).
\end{equation}
Upon rearranging this equation, the large scale Lagrange halo bias $b_L$ for halos of mass $M$ arising from Gaussian initial conditions is related to the halo number density as 
\begin{equation}
b_L^G = \bar{n}^{-1} \frac{\partial n}{\partial \delta_l} = \bar{n}^{-1} \frac{\partial n}{\partial \delta_c},
\end{equation}
because in the Gaussian case the effect of modulating the density field by a long wavelength mode $\delta_l$ in some volume of the Universe can be rexpressed simply as an additive change in the effective critical density for collapse $\delta_c$ in that region.  In the presence of mode coupling due to primordial non-Gaussianity, large-scale modes affect the statistical properties of small-scale modes (and vice-versa). For special types of non-Gaussianity (i.e. the local case) it is possible to generalize the Gaussian peak-background split  derivation of halo clustering to non-Gaussian initial conditions. This was done in Slosar et al.~(2008) \cite{slosar/etal:2008}, which we now summarize.
\subsection{Review of Slosar et al.~(2008) theoretical results}
Slosar et al.~(2008) \cite{slosar/etal:2008} re-cast previous results on non-Gaussian halo bias \cite{dalal/etal:2008b,matarrese/verde:2008,afshordi/tolley:2008} by instead using a peak-background split of the Gaussian field $\phi$ in Equation \ref{eq:fnlnongauss}, where long and short wavelength modes are independent:
\begin{equation}
\label{eq:peakbkgd}
\phi = \phi_{l} + \phi_{s}.
\end{equation}
The long wavelength density and potential fluctuations are related by the Poisson equation, which can be expressed in Fourier space using $\delta_l(k,z_0) = D(z_0) {\cal M}(k) \Phi(k)$ with
\begin{equation}
{\cal M}(k) = \frac{2 c^2 k^2 T(k)}{3\Omega_m H_0^2},
\end{equation}
where $T(k)$ is the transfer function and $D(z) = g(z) (1+z)^{-1}$ is the linear growth function normalized to $(1+z)^{-1}$ in the matter-dominated epoch.  That is, $g(z)$ is the growth suppression due to non-zero $\Lambda$, for which $g(z_{CMB}) = 1$ and $g(z=0) \approx 0.75$ in a concordance cosmology.  We note here that while in this paper $\Phi(k)$ refers to the potential in the matter dominated epoch, other authors (e.g., Ref.~\cite{grossi/etal:2009}) have chosen to work with $\tilde{\Phi}(k)$ normalized at $z=0$, the ``LSS convention.''  The gravitational potential depends on redshift in a non Einstein-de Sitter universe: $\tilde{\Phi}(k) = \Phi(k) g(z=0)/g(z_{CMB})$.  We can see from Equation \ref{eq:fnlnongauss} that $f_\mathrm{NL}$ also depends on this choice, so that $f^{LSS}_\mathrm{NL} =  f^{CMB}_\mathrm{NL} g(z_{CMB})/g(z=0) \approx 1.3 f^{CMB}_\mathrm{NL}$.  Throughout this section, $f_\mathrm{NL}$ refers to $f^{CMB}_\mathrm{NL}$.

The effect of the non-Gaussianity described by Equation \ref{eq:fnlnongauss} (and its induced mode coupling) is to modulate the amplitude of small-scale density fluctuations $\delta_s$ with the long wavelength potential fluctuations.  This can be viewed as a change in the local value of $\sigma_8$,  $\sigma_8^\mathrm{local}$, due to $\phi_l$:
\begin{equation}
\delta_s(z_o) = D(z_o) {\cal M}(k) \left[(1+ 2f_\mathrm{NL} \phi_l) \phi_s+f_\mathrm{NL} \phi_s^2\right].
\end{equation}
In this picture, the Lagrangian halo bias becomes
\begin{equation}
\label{eq:scaledepbias}
b_L(M,k,z_o) = b_L^{G}(M, z_0) + 2 f_\mathrm{NL} \frac{d\phi_l(k)}{d\delta_l(k, z_o)}\frac{\partial \; \mathrm{ln} \; n}{\partial \; \mathrm{ln} \; \sigma_8^\mathrm{local}}.
\end{equation}
The final expression for the non-Gaussian scale-dependent component of the halo bias on large scale is
\begin{equation}
\Delta b_\mathrm{NG}(M,k,z_o) = \frac{2 f_\mathrm{NL}}{D(z_o) {\cal M}(k)} \frac{\partial \; \mathrm{ln} \; n(M,z_o)}{\partial \; \mathrm{ln} \; \sigma_8},
\end{equation}
where we have dropped the ``local'' label.  That is, under the assumptions of \cite{slosar/etal:2008}, the non-Gaussian scale-dependent halo bias can be predicted from determination of the dependence on $\sigma_8$ of the mass function of the desired objects in cosmologies with Gaussian initial conditions.  In particular, to determine $\Delta b_\mathrm{NG}$ for halos of mass $M_0$ that have undergone a recent merger, they write
\begin{equation}
\frac{\partial \; \mathrm{ln} \; n_\mathrm{merger}(M_o, z_o)}{\partial \; \mathrm{ln} \; \sigma_8} = \frac{\partial \; \mathrm{ln} \; n(M_o, z_o)}{\partial \; \mathrm{ln} \; \sigma_8} + \frac{\partial \; \mathrm{ln} \; P(M_1 | M_o, z_o)}{\partial \; \mathrm{ln} \; \sigma_8}
\end{equation}
where $M_1$ is the progenitor mass for a halo of mass $M_0$ that has undergone a recent merger, and $P(M_1 | M_o, z_o)$ is the probability that a halo of mass $M_o$ at $z_o$ has a recent progenitor of mass $M_1$.  For a universal mass function, the first term evaluates to $\delta_c b_L^G$ (or $q\delta_c b_L^G$ \cite{grossi/etal:2009}).  Using the extended Press-Schechter (ePS) formalism, the second term evaluates to -1 (independent of $M_1$), in good agreement with the dependence of merger rates on $\sigma_8$ found in $N$-body simulations during the matter-dominated epoch \cite{slosar/etal:2008}.  
Ref. \cite{furlanetto/kamionkowski:2006}  finds that in the  ePS formalism, the Gaussian halo bias  is independent of its formation history. While ``halo assembly bias"  is the subject of a lot of current theoretical effort (e.g.,\cite{gao/white:2007, dalal/etal:2008} and references therein) it appears to be substantial only at the lowest halo masses.
$N$-body simulations show that the dependence of halo bias on secondary paremeters is relatively small for the massive halos on very large scales of interest in this work \cite{gao/springel/white:2005,wechsler/etal:2006, wetzel/etal:2007, jing/suto/mo:2007, dalal/etal:2008, faltenbacher/white:2010}.  
In particular, Ref.~\cite{jing/suto/mo:2007} find that for $M \geq 10 M_{\star}$, the 20\% youngest halos have a 10\% larger Gaussian halo bias than the 20\% older halos.
Here we concentrate instead on the formation history dependence of the non-Gaussian correction to the halo bias, which, as we show below, is much larger.

\subsection{General dependence of $\Delta b_\mathrm{NG}$ on $z_f$ using ePS}
In the formalism of ePS, we can easily generalize the results of \cite{slosar/etal:2008} to express the dependence of $\Delta b_\mathrm{NG}$ on the halo formation history, which we define by the ``formation redshift'' $z_f$.  In the original formulation of ePS  \cite{lacey/cole:1993}, $z_f$ is the redshift at which the halo contains half of its current mass;  on the other hand Ref.~\cite{viana/liddle:1996} suggests that, at least for the observational properties of clusters, this should rather be defined as $M_f = f M_0$, with $f \approx 0.75$.  In this section  we leave $f$ as a free parameter.  We can therefore write $\Delta b_\mathrm{NG}$ explicitly in terms of the halo mass $M$ observed at redshift $z_0$ with formation redshift $z_f$, and fraction of the halo mass at $z_f$, $f$.
\begin{eqnarray}
\Delta b_\mathrm{NG}(M,k,z_o, z_f, f) = \nonumber \\
\frac{2 f_\mathrm{NL}}{D(z_o) {\cal M}(k)} \left(\frac{\partial \; \mathrm{ln} \; n(M, z_o)}{\partial \; \mathrm{ln} \; \sigma_8} + \frac{\partial \; \mathrm{ln} \; P_{z_f}(f M, z_f | M, z_0)}{\partial \; \mathrm{ln} \; \sigma_8}\right) \label{myNGbias}
\end{eqnarray}
Here $P_{z_f}$ is the conditional mass function -- the probability that a halo with mass $M$ at $z_o$ has a mass $f M$ at an earlier redshift between $z_f$ and $z_f + dz_f$.  With the same approximations as in Sec 2.5.2 of \cite{lacey/cole:1993}, but generalizing to arbitrary fraction $f$ defining the epoch of formation, an analytic expression for the probability distribution of formation redshifts $P_{z_f}(f M, z_f | M, z_o)$ can be derived as follows. 
We start by defining the amplitude of fluctuations in the linear density field evolved to $z=0$, as usual:
\begin{eqnarray}
\sigma^2(M) = \frac{1}{2\pi^2} \int {\cal P}(k) k^2 \hat{W}_{M}^2(k) dk = \sigma_8^2 \frac{\int {\cal P}(k) k^2 \hat{W}_{M}^2(k) dk}{\int {\cal P}(k) k^2 \hat{W}_{R = 8}^2(k) dk} \label{eq:sigmaofM}
\end{eqnarray}
where $\hat{W}_M$ is the Fourier transform of a top-hat filter with radius $R= (3M/4\pi\bar{\rho}_{m})^{1/3}$, $\bar{\rho}_{m}$ denotes the matter background density at $z=0$, and ${\cal P}(k)$ is the linear matter power spectrum at $z=0$.  The second equality makes explicit how we define the $\sigma_8$ in Equation \ref{myNGbias} that we differentiate with respect to.  That is, $\sigma_8$ defines the amplitude of $\sigma^2(M)$, while the mass dependence is held fixed.
Following Ref \cite{lacey/cole:1993} we also introduce the quantity 
 \begin{eqnarray}
\tilde{\omega}_f = \frac{\delta_c(z_f) - \delta(z_0)}{\sqrt{\sigma^2(f M) - \sigma^2(M)}}\,.\label{wtildedef}
\end{eqnarray}
At fixed final halo mass $M$ at $z_o$, the probability that the halo formed between $z_f$ and $z_f + dz_f$ is simply expressed in terms of $\tilde{\omega}_f$:
\begin{eqnarray}
P_{z_f} dz_f = P_{\tilde{\omega}_f} \frac{d\tilde{\omega}_f}{dz_f} dz_f = \nonumber \\
\left[\sqrt{\frac{2}{\pi}} \left(\frac{1}{f} - 2\right) e^{-\tilde{\omega}_f^2/2} -2\tilde{\omega}_f \left(\frac{1}{f} - 1\right){\rm Erfc}(\tilde{\omega}_f/\sqrt{2})\right] \frac{d\tilde{\omega}_f}{dz_f} dz_f  \label{dPdweqn}
\end{eqnarray}
where $\delta_c(z) = \Delta_c(z) D(z)/D(z=0)$ (with $\Delta_c(z) \approx 1.686$) is the critical overdensity for collapse at redshift $z$ and ${\rm Erfc}$ denotes the complementary error function.  The halo mass, redshift, and $\sigma_8$ dependencies of $P_{z_f}$ are absorbed into the variable $\tilde{\omega}_f$.  To compute the second term in Equation \ref{myNGbias}, we must differentiate the formation redshift probability distribution $P_{z_f}$ at fixed halo mass $M$ with respect to $\sigma_8$:
\begin{equation}
\label{eq:finalderiv}
\frac{\partial \; \mathrm{ln} \; P_{z_f}(f M,z_f | M,z_o)}{\partial \; \mathrm{ln} \; \sigma_8} = -1 - \frac{\tilde{\omega}_f}{P_{\tilde{\omega}_f}}\frac{dP_{\tilde{\omega}_f}}{d\tilde{\omega}_f}.
\end{equation}
Therefore, the ePS formalism predicts that the amplitude of the merger history dependent contribution to $\Delta b_\mathrm{NG}$ depends on $M$, $z_o$, and $z_f$ only through a single variable, $\tilde{\omega}_f$.  Note that Equation \ref{eq:finalderiv} approaches ${\tilde \omega}_f^2-1$ in the limit of large ${\tilde \omega}_f$.  In Section \ref{simresults}, we test Equation \ref{eq:finalderiv} explicitly by dividing our simulated halo sample into bins in $\tilde{\omega}_f$.  Correspondingly, we must average Equation \ref{eq:finalderiv} over the same bins.  To reduce the impact of known discrepancies between the ePS prediction for $P_{z_f}$ and those measured in $N$-body simulations (e.g., \cite{vandenbosch:2002}), we express the ePS predictions in terms of the fraction $x$ of halos with the lowest or highest values of $\tilde{\omega}_f$.  For the lowest, we set $\tilde{\omega}_1=0$ and solve for $\tilde{\omega}_2$ such that
\begin{equation}
\int_{0}^{\tilde{\omega}_2}d\tilde{\omega}_f P_{\tilde{\omega}_f} = x \; .
\end{equation}
For the highest values of $\tilde{\omega}_f$, we set $\tilde{\omega}_2=\infty$ and we solve similarly for $\tilde{\omega}_1$.  The final expression for the  mean value  of Equation \ref{eq:finalderiv} in the range [$\tilde{\omega}_{1}(x)$, $\tilde{\omega}_{2}(x)$] as a function of halo fraction $x$ is
\begin{eqnarray}
\left <\frac{\partial \; \mathrm{ln} \; P_{z_f}(f M, z_f | M, z_o)}{\partial \; \mathrm{ln} \; \sigma_8}\right> =  \label{eq:finalderivavg} \\
\frac{\int_{\tilde{\omega}_1(x)}^{\tilde{\omega}_2(x)}d\tilde{\omega}_f P_{\tilde{\omega}_f} \left(-1 - \frac{\tilde{\omega}_f}{P_{\tilde{\omega}_f}}\frac{dP_{\tilde{\omega}_f}}{d\tilde{\omega}_f}\right)}{\int_{\tilde{\omega}_1(x)}^{\tilde{\omega}_2(x)}d\tilde{\omega}_f P_{\tilde{\omega}_f}} 
= \frac{\left[-\tilde{\omega}_f P_{\tilde{\omega}_f}\right]_{\tilde{\omega}_1(x)}^{\tilde{\omega}_2(x)}}{\int_{\tilde{\omega}_1(x)}^{\tilde{\omega}_2(x)}d\tilde{\omega}_f P_{\tilde{\omega}_f}}\,. \nonumber
\end{eqnarray}
One may worry that the expression for the conditional probability (Equation \ref{dPdweqn}) was derived within the Press-Schecter \cite{press/schechter:1974} framework: an approximation yielding an expression for the halo mass function which does not reproduce well N-body simulation results especially at small and big masses and  which has been significantly improved (e.g., \cite{sheth/mo/tormen:2001, jenkins/etal:2001}).  Unfortunately there is no analytical expression for the conditional probability in the context of these improvements.  However, van den Bosch et al. \cite{vandenbosch:2002} found relatively good agreement between Equation \ref{dPdweqn} and their $N$-body simulation results, though for the massive halos of interest in this work, the simulated halos formed somewhat earlier than Equation \ref{dPdweqn} predicts.  Moreover, here we are only interested in how $dP/dz_f$ changes as $\sigma_8$ is varied; we therefore  expect ePS to fare better in this respect. We will show in section \label{simresults} that the ePS approach adopted here is a remarkably good description of the dependence of $dP/dz_f$ on $\sigma_8$ as measured from N-body simulations.
\section{Simulation Results}
\label{simresults}
\begin{figure}
  \centering
 \includegraphics[scale=0.5, angle=270]{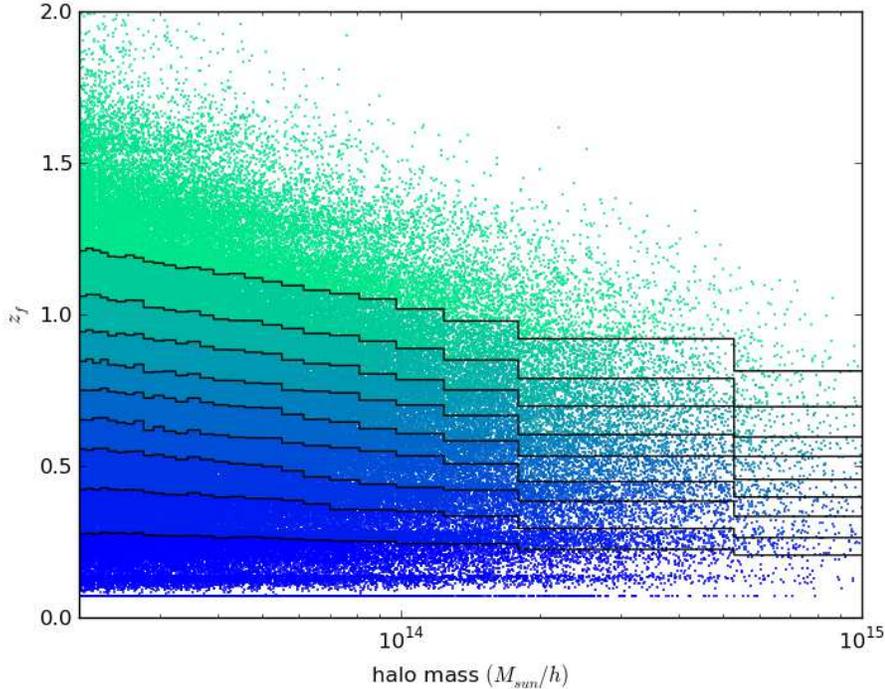}
  \caption{\label{fig:Mvszf}  Each point in the figure represents a halo in our mass-limited $M \geq 2 \times 10^{13} h^{-1} M_{\odot}$ sample at $z=0$.  Each color band represents a subsample of the full distribution that reproduces the mass function of the full sample but has different formation redshift distributions; each subsample contains 10\% of the total halos in the sample.  The black lines dividing two colors define our cumulative samples.  For instance, all halos below the third black line from the bottom of the plot enter our $x=0.3$ ``low $z_f$'' sample, while all halos above the third black line from the top enter our $x=0.3$ ``high $z_f$'' sample.}
\end{figure}
We use the Grossi et al.~(2009) \cite{grossi/etal:2009} set of 5 simulations ($f_\mathrm{NL}^\mathrm{LSS} = 0, \pm 100, \pm 200$, or equivalently, $f_\mathrm{NL}^\mathrm{CMB} = 0, \pm 75, \pm 151$), where all simulations use the same initial condition field $\phi$ in Equation \ref{eq:fnlnongauss} to suppress cosmic variance.  These simulations have $L_\mathrm{box} = 1200 \; h^{-1}$ Mpc and particle mass $m_p = 1.4\times 10^{11} \; h^{-1} M_{\odot}$. We generate halo merger trees at $z=0$ with the SUBFIND code \cite{springel/etal:2005}, based on simulation outputs at $z=$ (0, 0.137, 0.283, 0.441, 0.613, 0.804, 1.017, 1.258, 1.535, 1.857, 2.236, 2.688, 3.235, 3.907, 4.749, 5.822).  To define a halo's formation redshift $z_f$, we interpolate between the progenitor masses at the available redshifts.  

The first term in Equation \ref{myNGbias} is proportional to the Gaussian Lagrange halo bias $b_L^G$, which depends on halo mass.  To separate the dependence of $\Delta b_\mathrm{NG}$ on halo formation redshift from its dependence on halo mass, we create $z_f$-dependent subsamples of the full mass-limited halo sample at fixed observed redshift $z_o$ that match the halo mass function of the full sample.  To do this, we sort the $N_\mathrm{halos}$ halos by mass, and form groups of $N_\mathrm{group} \ll N_\mathrm{halos}$ halos closest in mass.  We sort these $N_\mathrm{group}$ halos by $z_f$, or equivalently by $\tilde{\omega}_f$ (since the halo mass is nearly constant in the $N_\mathrm{group}$ halo subsample).  The highest and lowest fraction $x$ of these $N_\mathrm{group}$ halos enter the $z_f$-dependent subsamples that by design have matching mass functions.  The full $M \geq 2 \times 10^{13} h^{-1} M_{\odot}$ sample has $N_\mathrm{halos} \approx 250,000$ at $z=0$, while we use $N_\mathrm{group} = 10,000$ throughout the main text; in the Appendix we demonstrate that our results are unchanged if $N_\mathrm{group} = 100$ is used instead.  Figure \ref{fig:Mvszf} illustrates the sample selection scheme more clearly, where we plot each halo in our $z=0$, $M \geq 2 \times 10^{13} h^{-1} M_{\odot}$ sample in the two dimensional space $z_f$-$M$, with the formation in this case defined by $M(z_f) = f M_o$ for $f=0.5$.  Note the relatively mild trend that the average $z_f$ decreases with $M$, and also the large spread in $z_f$ at fixed $M$.  At each small halo mass bin defined by $N_\mathrm{group}$ halos, we divide the sample into 10 bins based on $z_f$, and the $z_f$ bins from different mass bins are combined to produce halo samples with matching mass functions but different formation redshift distributions.  The result is 10 distinct samples represented by the color bands in Figure \ref{fig:Mvszf}.  To increase the signal to noise ratio of our measurements, we do not present non-Gaussian halo bias measurements for the 10 disjoint samples represented in Figure \ref{fig:Mvszf}, but instead present results for cumulative samples defined by the 9 lines which each divide two colors in the figure.  That is, samples labelled as $x = 0.3$ contain either the three lowest or three highest color bands in the figure; the black lines divide the different samples.

\begin{figure}
  \centering
 \includegraphics[scale=0.55]{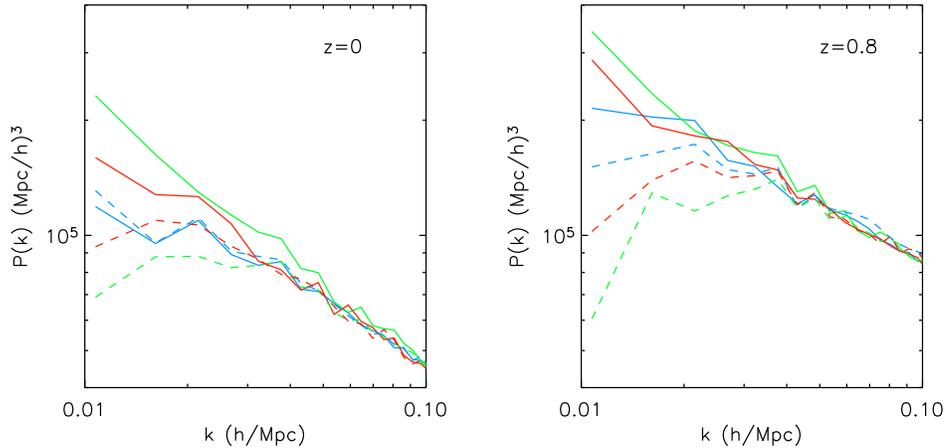}
  \caption{\label{fig:ngspectra1}  Halo-halo power spectra from the $f_\mathrm{NL}^\mathrm{LSS} = -200$ (dashed) and $f_\mathrm{NL}^\mathrm{LSS} = 200$ (solid) simulations for halos with $M \geq 2 \times 10^{13} h^{-1} M_{\odot}$ for 25\% subsamples of halos with the highest (green) and lowest (blue) $z_f$ values.  No shot noise correction has been applied, and so for comparison we produce a random subsample of the full mass-limited sample with the same number density as the other samples (red).  The left panel uses halos at $z=0$, while the right panel uses halos at $z=0.8$.}
\end{figure}

Figure \ref{fig:ngspectra1} illustrates the signal we quantify in this section.  We plot the halo-halo power spectrum for halos with $M \geq 2 \times 10^{13} h^{-1} M_{\odot}$ at $z=0$ (left panel) and $z=0.8$ (right panel) for two subsamples of the parent sample with the 25\% highest (blue) and lowest (green) formation redshifts $z_f$ for $f=0.5$, selected as described in the previous paragraph.  For reference we plot a random subsample of the halo population with the same number density as the $z_f$-dependent subsamples in red. This should have approximately the same noise properties as the other two samples, but reflects the clustering properties of the full halo population. Dashed lines are for $f_\mathrm{NL}^\mathrm{LSS}=-200$ and solid lines are for    $f_\mathrm{NL}^\mathrm{LSS}=200$. In the region of the spectrum where $\Delta b_\mathrm{NG}$ is important (restricted to $k \leq 0.03 \; h/$Mpc in our analysis), the power spectrum is noisy but appears to depend on $z_f$.  Because the spectra show good agreement at higher $k$ where $\Delta b_\mathrm{NG}$ is small, our scheme to match the first term in Equation \ref{myNGbias} between samples is successful, and the difference between the spectra at small $k$ can be attributed to the second ($z_f$-dependent) term in Equation \ref{myNGbias}, rather than the term proportional to the Gaussian Lagrange bias $b^L_G$.

We assume the following relation between  halo ($\delta_h({\bf k})$) and dark matter ($\delta_m({\bf k})$) individual Eulerian density modes observed at redshift $z_o$, in order to fit for the amplitude of the non-Gaussian halo bias, $A_\mathrm{NG}$:
\begin{equation}
\label{deltahmodel}
\delta_h({\bf k}) = \left(b_G + A_\mathrm{NG} \frac{2 f_\mathrm{NL}}{D(z_o) {\cal M}(k)}\right) \delta_m({\bf k}) + n({\bf k}).
\end{equation}
Here $b_G = 1 + b_L^G$ represents the Eulerian scale-independent contribution to the halo bias (first term in Equation \ref{eq:scaledepbias}).  $A_\mathrm{NG}$ describes the amplitude of the non-Gaussian halo bias, and Equation \ref{myNGbias} predicts that $A_\mathrm{NG}$ is the logarithmic derivative of the Gaussian mass function of the halo sample with respect to $\sigma_8$, i.e., $A_\mathrm{NG}$ should be given by $\partial \; \mathrm{ln} \; n(M, z_o)/\partial \; \mathrm{ln} \; \sigma_8 + \partial \; \mathrm{ln} \; P_{z_f}(f M, z_f | M, z_0)/\partial \; \mathrm{ln} \; \sigma_8$.  We assume the noise $n({\bf k})$ to be Poissonian.  For the results presented in this section, we first use the $f_\mathrm{NL}^{LSS} = 0$ simulation to determine the scale-independent halo bias $b_G$ for each mass, redshift, and $z_f$-dependent halo subsample while holding $A_\mathrm{NG} = 0$.  We then assume the $f_\mathrm{NL}$ and $k$ dependence in Equation \ref{deltahmodel}, and fit the four simulations with non-zero $f_\mathrm{NL}$ for a single number, $A_\mathrm{NG}$, for each halo subsample.  This accounts for any small dependence of $b_G$ on $z_f$ in the Gaussian case.  In the Appendix we present further details of this fitting procedure and compare this approach with a more conservative one, where separate values of $b_{G}$ are fit simultaneously for each $f_\mathrm{NL} \neq 0$ when fitting for $A_\mathrm{NG}$.  An advantage over previous approaches is that our fit is performed mode by mode for the 366 available modes, rather than to a power spectrum where an effective value of $k$ must be chosen for each bin; this choice seemed to impact the results of Ref.~\cite{pillepich/porciani/hahn:2010}.  Note that in the model given by Equation \ref{deltahmodel}, $\sigma^2_{A_\mathrm{NG}} \propto 1/N_\mathrm{halos}$, so our measurement error is smallest at $z=0$ where there are the most halos above our fixed mass limit available from our simulations.

\begin{figure}
\centering
\resizebox{1\columnwidth}{!}{\includegraphics{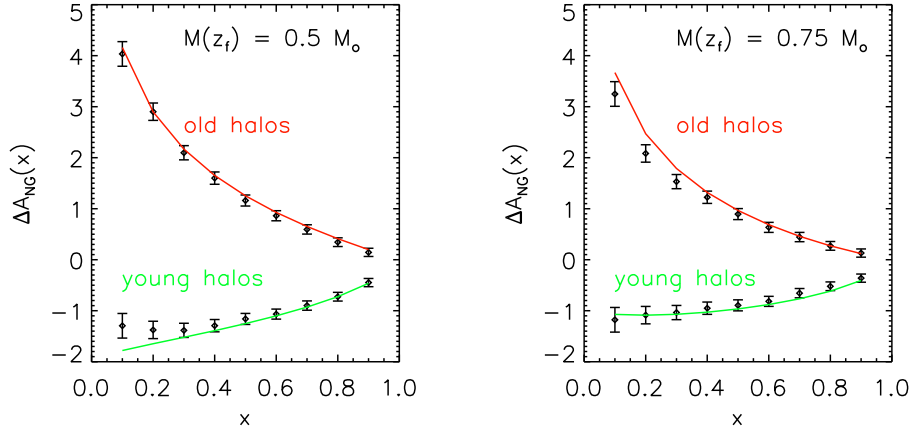}}
\caption{\label{fig:simsvsePS}  $\Delta A_\mathrm{NG}^\mathrm{h/l}(x)  = A_\mathrm{NG}^\mathrm{h/l}(x) - A_\mathrm{NG}^\mathrm{all}$ averaged over simulation snapshots between $z=0$ and $z=2.23$ for subsamples of $M \geq 2 \times 10^{13} \; h^{-1} \; M_{\odot}$ halos of the highest (lowest) fraction $x$ of halo formation redshifts (black points with errors).  The left panel uses $f=0.5$ to define the formation redshift $z_f$, while the right panel uses $f=0.75$.  The ePS prediction (Equation \ref{eq:finalderivavg}) for the high $z_f$ (red) and low $z_f$ (green) subsamples provides an excellent fit to the simulation measurements.}
\end{figure}

Equation \ref{eq:finalderivavg} predicts that, when written as a function of the variable $x$,
the $z_f$-dependent fractional contribution to $\Delta b_\mathrm{NG}$ is independent of both halo mass $M$ and observed redshift $z_o$; we will check these predictions explicitly later in this section.  We begin with a mass-limited halo sample $M \geq 2\times 10^{13} \; h^{-1} \; M_{\odot}$ at each snapshot redshift $z_o$ available from our simulations and measure the amplitude of its non-Gaussian bias term, $A_\mathrm{NG}^\mathrm{all}(z_o)$.  To quantify the dependence of $A_\mathrm{NG}$ on $z_f$, we measure $\Delta A_\mathrm{NG}^\mathrm{h/l}(x, z_o)$:
\begin{equation}
\Delta A_\mathrm{NG}^\mathrm{h/l}(x,z_o) = A_\mathrm{NG}^\mathrm{h/l}(x, z_o) - A_\mathrm{NG}^\mathrm{all}(z_o),
\end{equation}   
where $x$ is the fraction of halos with the highest (lowest) formation redshifts entering the $h$ ($l$) subsamples at each $z_o$.  In Figure \ref{fig:simsvsePS} we plot $\Delta A_\mathrm{NG}^\mathrm{h/l}(x)$ after performing an error-weighted average over all $z_o$ values between $z=0$ and $z=2.23$ for two definitions of the halo formation redshift, $f=0.5$ and $f=0.75$.  Note that $\Delta A_\mathrm{NG}^\mathrm{h}(x)$ is positive, while $\Delta A_\mathrm{NG}^\mathrm{l}(x)$ is negative for both values of $f$.  Conservatively, we show the errors from the $z_o=0$ sample only, since the halo samples at different redshifts will be correlated.  Moreover, note that for two values of $x$, $x_1 < x_2$, the first subsample is contained in the second.  Therefore, the error bars at different $x$ are highly correlated as well.
The agreement with the ePS prediction given by Equation \ref{eq:finalderivavg} is excellent for both the low $z_f$ (green curve) and high $z_f$ (red curve) subsamples.  Note that the ePS prediction is:
\begin{equation}
\Delta A_\mathrm{NG}=\frac{\partial \; \mathrm{ln} \; P_{z_f}(f M,z_f | M,z_o)}{\partial \; \mathrm{ln} \; \sigma_8} =-1- \frac{\tilde{\omega}_f}{P_{\tilde{\omega}_f}}\frac{dP_{\tilde{\omega}_f}}{d\tilde{\omega}_f}.
\end{equation}
For the finite binning used in the simulations, the plotted theory line uses Equation \ref{eq:finalderivavg}.

\begin{figure}
\centering
\resizebox{1.03\columnwidth}{!}{\includegraphics{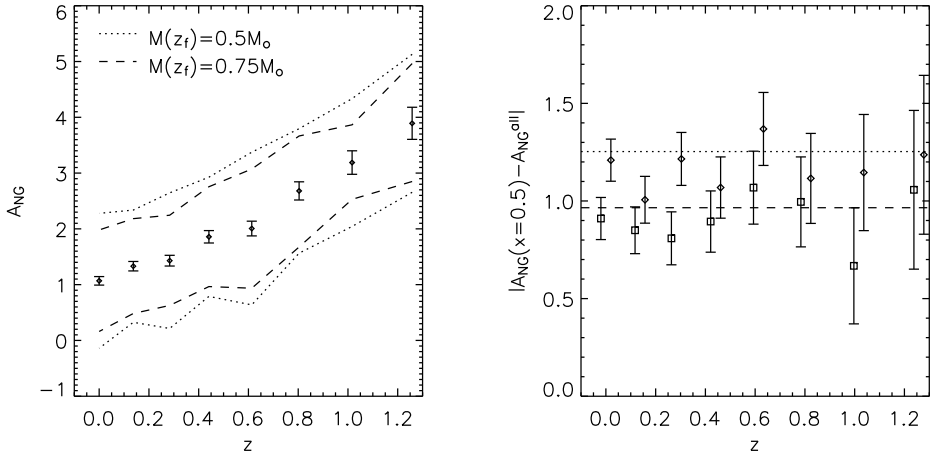}}
\caption{\label{fig:zindep}  {\em Left panel:} Best fit $A_\mathrm{NG}$ for the full mass-limited sample $M \geq 2 \times 10^{13} \; h^{-1} \; M_{\odot}$ (points with errors) as a function of redshift, as well as for subsamples split in half on $z_f$ ($x=0.5$) for $f=0.5$ (dotted) and $f=0.75$ (dashed).  The $z_f$-dependent correction is well described as an additive correction as in Equation \ref{myNGbias}.  {\em Right panel:} $|A^\mathrm{h/l}_\mathrm{NG}(x=0.5) - A_\mathrm{NG}^\mathrm{all}|$ as a function of redshift for $f=0.5$ (diamonds) and $f=0.75$ (squares).  Points are offset from the snapshot redshift by $\pm 0.02$ for clarity, and the ePS predictions are shown as a straight lines for $f=0.5$ (dotted) and $f=0.75$ (dashed).  There is no clear redshift dependence, though the errors are large so our measurement is not very constraining.}
\end{figure}

\begin{figure}
\centering
\resizebox{\columnwidth}{!}{\includegraphics{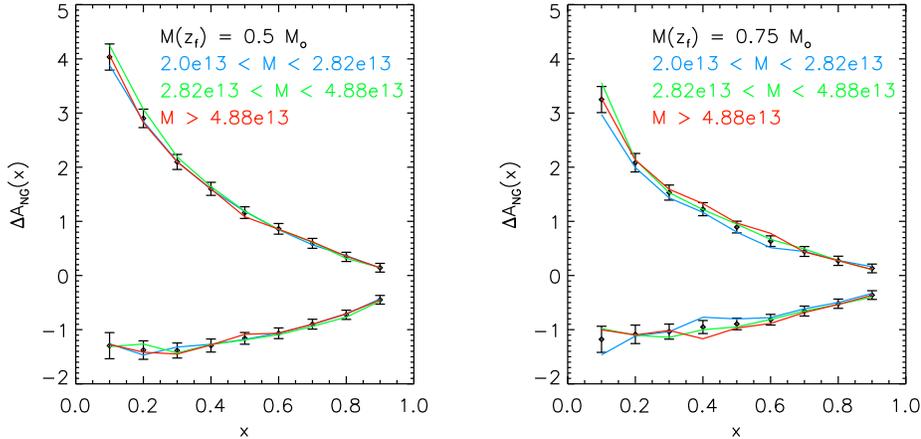}}
\caption{\label{fig:massindep}  The average profile measured using all halos with $M \geq 2 \times 10^{13} \; h^{-1} \; M_{\odot}$ (black points with error bars, same as in Figure \ref{fig:simsvsePS}) plotted with three subsamples in mass containing equal numbers of halos.  The blue curve uses halos with $M \in [2.0, 2.82] \times 10^{13} \; h^{-1} \; M_{\odot}$, the green curve uses halos with $M \in [2.82, 4.88] \times 10^{13} \; h^{-1} \; M_{\odot}$, and the red curve uses halos with $M \geq 4.88 \times 10^{13} \; h^{-1} \; M_{\odot}$. For the mass range probed here, $\Delta A_\mathrm{NG}$ is independent of mass.}
\end{figure}

The essential features of the non-Gaussian halo assembly bias we quantify here are:
\begin{itemize}
\item The left panel of Figure \ref{fig:zindep} shows how the relative importance of the first and second term in Equation \ref{myNGbias} evolves with redshift, for our {\em mass-limited} halo sample; the amplitude of the $z_f$-dependent contribution is potentially large compared to the first term in Equation \ref{myNGbias}, $\partial \; \mathrm{ln} \; n(M, z_o)/\partial \; \mathrm{ln} \; \sigma_8 \sim 1.68 (b_G - 1)$.  For instance, if we split the $z=0$, $M \geq 2\times 10^{13} \; h^{-1} \; M_{\odot}$ halo sample in two as a function of $z_f$ (i.e., $x=0.5$), the $A_\mathrm{NG}$ predictions differ from the mean ($A_\mathrm{NG}^\mathrm{all}=1.12$) by $\pm (1.28 \pm 0.2)$.  That is, the more recently formed halos have a factor of $\sim 7$ smaller expected signal than the full halo sample (and with opposite sign: $A_\mathrm{NG}^\mathrm{l}/A_\mathrm{NG}^\mathrm{all}\simeq -0.16/1.12$), while the older halos have a factor of $\sim 2$ larger signal than the full halo sample ($A_\mathrm{NG}^\mathrm{h}/A_\mathrm{NG}^\mathrm{all}\simeq 2.3/1.12$).  
\item The effect is asymmetric between old and young halos.  Even if the tracer population excludes only the 10\% oldest halos, the value of $A_\mathrm{NG}$ for the remaining 90\% of the halos differs from the full sample by $\approx 0.44$; at $z=0$, this amounts to a change of $0.44/1.12 = 40\%$.
\item Figures \ref{fig:zindep} and \ref{fig:massindep} support the ePS prediction expressed in Equation \ref{eq:finalderivavg} that there is no mass or redshift dependence to the $z_f$-dependent term, when expressed in terms of the variable $x$.  However, note that we are restricted to studying massive halos $M \geq 2 \times 10^{13} \; h^{-1} \; M_{\odot}$, and that our errors on $\Delta A_\mathrm{NG}$  rapidly increase with $z$.
\end{itemize}
Finally, we generate the closest possible sample to recent major mergers available from our simulations at $z=0$, where the errors on $\Delta A_\mathrm{NG}$ are smallest.  We sort the halo sample not on $z_f$ but on the first progenitor's mass at the previous snapshot output, $z=0.137$ (1.7 Gyr earlier).  For $x \leq 0.4$ (i.e., for the 40\% of halos with the lowest progenitor masses) we find a $\Delta A_\mathrm{NG}(x)$ consistent with $-1$, the ePS prediction derived in Slosar et al. 2008 \cite{slosar/etal:2008}, and the limit of our ePS predictions for large values of $f$.

\section{Implications for upcoming galaxy surveys}
\label{implications}
Non-Gaussianity constraints achievable from forthcoming and planned surveys using halo bias are very promising \cite{carbone/verde/matarrese:2008,carbone/mena/verde:2010}; these forecasts are obtained  using the mean non-Gaussian halo bias relation and therefore assume that these surveys will select galaxies that provide a fair sample of the underlying dark matter halo population in the suitable mass range. However, if the survey selection  were to preferentially select galaxies occupying dark matter halos with $z_f$ lower than the mean or were to miss e.g., the 10\% of dark matter halos with the highest $z_f$, this, if unaccounted for,  would introduce  a significant bias in the measured $f_\mathrm{NL}$ parameter. As shown in Section 3, in principle, for some extreme cases, this bias could be as large as 40-100\% of the (mean) signal.

As alarming as this may seem, it is important to bear in mind that the results of the previous section only affect the predictions of $\Delta b_\mathrm{NG}$ for galaxy redshift surveys if the probability of a halo of hosting a galaxy in the survey sample depends on the formation history of the halo, {\em at fixed halo mass $M$}.  Within the halo-model framework \cite{cooray/sheth:2002}, the standard implementation of the halo occupation distribution approach describes the bias of specific galaxy types with respect to the underlying dark matter by assuming that the probability for a halo to host a galaxy depends only on the halo mass.  While for many galaxy types the impact of secondary parameters appears small (e.g., \cite{skibba/etal:2006, skibba/sheth:2009,tinker/etal:2007} and references therein), there are some indications of halo-assembly bias and evidence for secondary parameters \cite{mateus/jimenez/gaztanaga:2008,cooper/etal:2008,cooper/etal:2010}.  Semi-analytic models of galaxy formation are dependent on the entire dark-matter halo-merger history and can, in principle, include the full dependence of galaxy properties on secondary parameters.  Ref.~\cite{croton/gao/white:2007} assessed the impact of {\em Gaussian} halo assembly bias for Millenium simulation galaxy samples with two different magnitude cuts, and found changes to the Gaussian bias of $\lesssim 10\%$.  However, this change was not explained by the addition of any simple secondary parameter, like $z_{f=0.5}$ or halo concentration.  Therefore, by considering $z_f$ in what follows, we may be underestimating the possible signal from other properties of halo formation that correlate more closely with galaxy properties.  

In this section we use galaxies selected from the Bertone et al. (2007) \cite{bertone/delucia/thomas:2007} semi-analytic galaxy--formation model implemented on the Millenium simulation halo merger trees \cite{springel/etal:2005}.  In order to estimate the order of magnitude of this effect, we consider realistic but simply selected galaxy populations relevant to upcoming galaxy surveys, and ask whether these galaxies occupy dark matter halos with specific formation histories, in such a way to introduce a marked non-Gaussian halo assembly bias contribution.  Note that  the same procedure  can be followed to study the impact of halo assembly bias  on galaxy clustering in {\em Gaussian} initial conditions; in fact, in both cases,  we only need to quantify how different the formation redshift distribution of the halo population hosting the selected galaxies is from the full halo population, for a given halo mass $M$ (or small $dM$ around $M$).

\begin{figure}
\centering
\resizebox{\columnwidth}{!}{\includegraphics{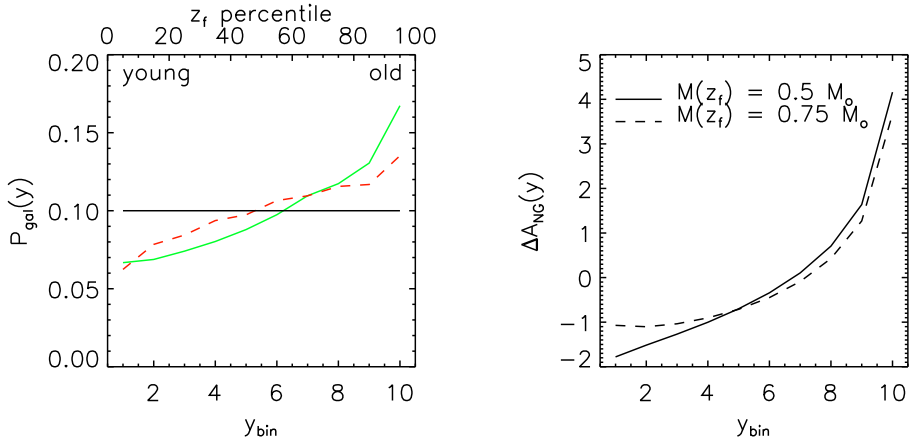}}
\caption{\label{fig:highmstellar1}  The solid (dashed) black curves in the right panel show $\Delta A_\mathrm{NG}(y)$ for 10 bins in $\tilde{\omega}_f$ for $f=0.5$ ($f=0.75$) using Equation \ref{eq:finalderivavg}.  Each bin contains 10\% of halos, in contrast to the theory curves in Figure \ref{fig:simsvsePS}, where the ``high'' and ``low'' curves considered cumulative samples (i.e., 30\% lowest and highest).  The solid green curve in the left panel shows $P_\mathrm{gal}(y)$ for the stellar mass selected sample described in the text (where $z_f$ has been defined using $f=0.5$), while the dashed red curve shows $P_\mathrm{gal}(y)$ for the star formation rate selected galaxies (where $z_f$ has been defined using $f=0.75$).  The normalization is chosen such that $P_\mathrm{gal}(y) = 0.1$ represents a sampling of the underlying halo distribution which is independent of halo formation redshift $z_f$.  The integral over the product of the solid (dashed) curves gives the assembly bias contribution to the non-Gaussian halo bias (Equation \ref{Anggal}) for the stellar mass (star formation rate) selected galaxy samples.  The result is $\Delta A_\mathrm{NG}^\mathrm{gal} = 0.51$ for the stellar mass selected sample, and $\Delta A_\mathrm{NG} = 0.24$ for the star formation rate selected sample.  See Table \ref{table:galaxystuff} for more details.}
\end{figure}

The procedure in the context of a mock galaxy sample embedded in an $N$-body simulation is as follows:
\begin{itemize}
\item Select a galaxy sample (e.g., based on observational selection criteria) and identify their host dark matter halos ({\it host halo sample}). 
\item Identify the host dark matter halos' mass range and select the full dark matter halo sample in that mass range ({\it full halo sample}). The halos containing the galaxy sample are a subset of this set.
\item Measure the formation-dependent quantity of interest, such as $z_f$, for  the full halo sample.
\item Determine the dependence of the $z_f$ distribution on halo mass $M$; in practice we do this exactly as for the sample shown in Figure \ref{fig:Mvszf}.  That is, we determine the dividing lines between 10 bins in $z_f$ (black lines in Figure \ref{fig:Mvszf}) as a function of halo mass.  
\item Determine how the halos containing the galaxy sample occupy these 10 bins, summarized by $P_\mathrm{gal}(y_\mathrm{bin})$; here we use $y_\mathrm{bin}$ to denote the bin number or transversal band in Figure \ref{fig:Mvszf}.  $y_\mathrm{bin} = 1$ corresponds to halos with 10\% lowest $z_f$ for each mass bin and the $y_\mathrm{bin} = 10$ corresponds to halos with the 10\% highest $z_f$ for each mass bin.
If galaxies were hosted in a random sample of the full halos, $P(y_\mathrm{bin})$ would be a constant.  Any deviation from a constant indicates a correlation between the galaxy selection and the host halo formation history.  We have chosen the normalization such that $\sum_\mathrm{bins} P(y_\mathrm{bin}) = 1$.
\item Determine the halo assembly bias correction factor. For our application this is $\Delta A_\mathrm{NG}^\mathrm{gal}$:
\begin{equation}
\label{Anggal}
\Delta A^\mathrm{gal}_\mathrm{NG} = \int \Delta A_\mathrm{NG}(y) P_\mathrm{gal}(y) dy \approx \sum_{i=1}^{10} \Delta A_\mathrm{NG}(y_i) P_\mathrm{gal}(y_i)\,.
\end{equation}
To evaluate $\Delta A^\mathrm{gal}_\mathrm{NG}$, in practice we break the halos into 10 bins as shown in Figure \ref{fig:Mvszf} and sum the expected signal, $\Delta A_\mathrm{NG}(y_i)$, over the fraction of galaxies in these discrete bins, $P_\mathrm{gal}(y_i)$.  We compute $\Delta A_\mathrm{NG}(y_i)$ using Equation \ref{eq:finalderivavg}.  For this application we are considering disjoint halo subsamples rather than the cumulative bins considered in Figure \ref{fig:simsvsePS}.  In the right panel of Figure \ref{fig:highmstellar1} we show the theory curves for 10 disjoint bins, where we have used $f=0.5$ ($f=0.75$) to define $z_f$ in the solid (dashed) curve.
\end{itemize}
If for some sample galaxies $\Delta A^\mathrm{gal}_\mathrm{NG}$ is not negligible compared to $A^\mathrm{all}_\mathrm{NG} \approx 1.68 (b_G - 1)$, then the effect of non-Gaussian assembly bias cannot be ignored. In particular recall that for recently formed halos, $\Delta A_\mathrm{NG}$ can even  cancel out $A_\mathrm{NG}$, erasing any non-Gaussian signature, if present. 

BigBOSS \cite{schlegel/etal:2009} plans to select luminous red galaxies out to $z\sim 1$ and emission-line galaxies at higher redshift; other proposed surveys (e.g., Euclid \cite{laureijs:2009}) will also target emission-line galaxies out to $z\sim 2$.  Emission-line galaxies are thought to have high star formation rates, possibly triggered by mergers.  Should galaxy mergers trace the host dark matter halo mergers, this selection effect could greatly reduce the expected signal for a given value of $f_\mathrm{NL}$.  Another large future survey suitable for this study is LSST; LSST will select all galaxies above a given magnitude cut, and thus its selection criterion should be less correlated to the host halos accretion history than the other two surveys. 

As an example, we consider two distinct galaxy samples from Ref.~\cite{bertone/delucia/thomas:2007} semi-analytic catalogs at $z=0.99$, one selected on large stellar mass ($M_\mathrm{stellar} > 8 \times 10^{10} \; h^{-1}\;M_{\odot}$) and the other on large star formation rate ($\dot{M}_\mathrm{stellar} > 24 M_{\odot}/\mathrm{yr}$), where both samples are chosen to have number densities of $4.5 \times 10^{-4} \; ({\rm Mpc}/h)^{-3}$, i.e., in the right ballpark for a BAO-focused survey.  $P_\mathrm{gal}(y)$ for these samples is shown by the solid green  and red dashed lines in Figure \ref{fig:highmstellar1} for stellar mass and star formation rate selected samples, respectively; the figure is normalized such that $P_\mathrm{gal}(y) = 0.1$ corresponds to a uniform sampling of the underlying halo distribution.  For both samples, galaxies occupy halos across the distribution of $z_f$, but with a preference for halos that formed early (high $y_\mathrm{bin}$), though the trend is stronger with stellar mass.  Using Eq. \ref{Anggal}, this preference translates to $\Delta A_\mathrm{NG}^\mathrm{gal} = 0.51$ for the stellar mass selected sample, while their Gaussian bias is $\approx 2.3$, as measured from their power spectrum on large scales.  For the star formation rate-selected sample, $\Delta A_\mathrm{NG}^\mathrm{gal} = 0.24$, while their bias is $\approx 1.3$.  Since $A_\mathrm{NG}^\mathrm{all} \approx \delta_c (b_G - 1)$, accounting for the non-Gaussian assembly bias amounts to a boost of  the expected scale-dependent halo bias of 23\% and 48\%, respectively.  Results for these galaxy samples are summarized in Table \ref{table:galaxystuff}.

\begin{table*}
\begin{center}
\begin{tabular}{ccccccc}
\multicolumn{6}{c}{Non-Gaussian assembly bias in the Bertone et al. (2007) mock galaxy catalogs}\\
\hline
$\bar{n} \; (h^{-1} {\rm Mpc})^{-3}$ & selection criteria & $\Delta A_\mathrm{NG}^\mathrm{gal}$ & $b_G$ & $\delta_c (b_G - 1) \approx A_\mathrm{NG}^\mathrm{tot}$ & \% change\\
\hline
$4.5 \times 10^{-4}$ & $\geq 8 \times 10^{10} \; h^{-1} \;M_{\odot}$ & 0.51 & 2.3 & 2.2 & 23\%\\
$4.5 \times 10^{-4}$ & $\geq 24 \; M_{\odot}/$yr & 0.24 & 1.3 & 0.51 & 48\%\\
\end{tabular}
\caption{\label{table:galaxystuff} We select two galaxy samples from the semi-analytic model of Bertone et al. (2007) \cite{bertone/delucia/thomas:2007}, which have been run on the Millenium simulation \cite{springel/etal:2005}.  Both samples are selected to have the same number density, and roughly the value that would be targeted for upcoming BAO surveys.  The first sample is selected based on stellar mass, and the second based on star formation rate.  Both galaxy samples preferentially occupy halos that formed earlier than average (see Figure \ref{fig:highmstellar1}), which translates into non-zero values of $\Delta A_\mathrm{NG}^\mathrm{gal}$.  We measure the Gaussian Eulerian bias $b_G$ from the galaxy clustering amplitude in the Millenium for both samples in order to infer the expected $A_\mathrm{NG}$ for each sample in the absence of non-Gaussian halo assembly bias.}
\end{center}
\end{table*}

\section{Conclusions}
We have demonstrated that the impact of assembly bias on the amplitude of the non-Gaussian halo bias can be quite strong.  We have expanded the arguments in Slosar et al. (2008) \cite{slosar/etal:2008} using extended Press-Schechter theory to express non-Gaussian assembly bias in terms of halo formation redshift $z_f$ for arbitrary $f \geq 0.5$, where $z_f$ is the redshift at which a halo has accreted a fraction $f$ of its final mass.  This theory predicts that halo subsamples containing a fraction $x$ of the earliest (latest) forming halos (compared with other halos with the same mass) have a non-Gaussian halo bias that differs from the full parent halo sample by a fractional correction dependent only on $x$; when using this variable, the non-Gaussian assembly bias correction is independent of halo mass and redshift.  The $N$-body simulations of Grossi et al. (2009) \cite{grossi/etal:2009} are in good agreement with these ePS predictions.

The implications of these results for galaxy redshift surveys are extremely uncertain.  If the commonly adopted assumption that the probability of a halo hosting a particular type of galaxy only depends on the halo mass, then there will be no non-Gaussian halo assembly bias contribution to the galaxy sample's non-Gaussian bias.  However, in principle, galaxy formation depends on the entire history of host dark matter halos to some degree, and semi-analytic models of galaxy formation attempt to account for this dependence.  In Section \ref{implications}, we found that a relatively mild preference for early-forming halos for both stellar mass and star formation rate selected $z=1$ samples translates into an increase in the expected non-Gaussian galaxy bias of $\sim 23-48\%$ compared with the average signal expected from the samples' Gaussian bias values.  This result is particularly counter-intuitive for star-forming galaxies, since star formation is triggered by galaxy mergers in these models.  One should bear in mind that a galaxy merger does not necessarily correspond to a major merger of the host dark matter halo, and that it is reasonable to expect some time-lag between the dark matter halo merger and the merger of the galaxies populating them.  Furthermore, we caution that we have not been extensive in our exploration of galaxy sample selection space, or precise enough to make predictions for upcoming experiments.  There may be certain populations for which this effect may be much larger or much smaller.  On the other hand, it is possible that in an analysis aimed at constraining $f_\mathrm{NL}$, one may be able to weight galaxies by some color or spectral property in order to enhance the non-Gaussian signal in the survey. More work is needed to further quantify the impact of such an approach on the recovered constraints  on $f_\mathrm{NL}$ from realistic surveys.
\section*{Acknowledgements}
We thank the anonymous referee for going beyond the call of duty and independently cross-checking our results.  Non-Gaussian $N$-body simulations have been performed on the IBM-SP5 at CINECA (Consorzio Interuniversitario del Nord-Est per il Calcolo Automatico), Bologna, with CPU time assigned under an INAF-CINECA grant, and on the IBM-SP4 machine at the ``Rechenzentrum der Max-Planck-Gesellschaft'' at the Max-Planck Institut f\"{u}r Plasmaphysik with CPU time assigned to the ``Max-Planck-Institut f\"{u}r Astrophysik'' and at the ``Leibniz-Rechenzentrum'' with CPU time assinged to the Project ``h0073.''  The Millennium Simulation databases used in this paper and the web application providing online access to them were constructed as part of the activities of the German Astrophysical Virtual Observatory.  BR is supported by OISE/0530095.  LV acknowledges support of FP7-PEOPLE-2007-4-3-IRG n. 20182, FP7-IDEAS Phys.LSS 240117 and MICCIN grant AYA 2008-03531. LV thanks  for hospitality IPMU, where the last stages of this work were carried out.  KD acknowledges the support of the DFG Priority Program 1177.  This research has been partially supported by ASI Contract No. I/016/07/0 COFIS and ASI/INAF Agreement I/072/09/0 for the Planck LFI Activity of Phase E2.  LM acknowledges contracts Euclid-Dune I/064/08/0, ASI-INAF I/023/05/0, and ASI INAF I/088/06/0.

\section*{References}
\bibliographystyle{JHEP}
\providecommand{\href}[2]{#2}\begingroup\raggedright\endgroup

\appendix
\section{Fitting non-Gaussian bias}
\label{fittingprocedure}
We wish to construct a $\chi^2$ to estimate the amplitude of halo assembly-bias in non-Gaussian simulations; in principle this would require knowledge of a 4-point function in the non-Gaussian theory.  We begin considering the expected errors in linear theory for Gaussian initial conditions, and linearly biased tracers that Poisson-sample the continuous matter density field.  Our model for the relation between the halo density field and the matter density field for each mode ${\bf k}$ is given by Equation \ref{deltahmodel}, and Poisson sampling implies $\left<n({\bf k}) n^{\star}({\bf k})\right> = \bar{n}^{-1}$.  We will keep the $k$-dependence of the non-Gaussian component of the bias fixed, ${\cal M}^{-1}(k)$, and fit for an amplitude of the non-Gaussian component, $A_\mathrm{NG}$, and scale-independent component, $b_G$.  Under these assumptions, the variance about the model is 
\begin{equation}
\label{halovar}
\!\!\!\!\!\left<\!\left(\delta_h^{\star} \delta_m\! -\!  \left(\!b_G + A_\mathrm{NG} \frac{2 f_\mathrm{NL}}{D(z_o) {\cal M}(k)}\!\right)\! \!\delta_m^{\star} \delta_m\!\right)\!^2\!\right>\! =\! \left<n({\bf k}) n^{\star}({\bf k})\right>\! \delta_m^{\star} \delta_m
\end{equation}
Here the average is over the Poisson noise, and the matter field is considered fixed.  For our $\chi^2$ calculation, we will only consider the real component of $\delta_h^{\star} \delta_m$.  Since the noise is assumed uncorrelated with $\delta_m$, the noise associated with only the real component is smaller by a factor of 2 than in Equation \ref{halovar}, which counts both real and imaginary components.
\begin{equation}
\label{chi2eq}
\chi^2 = \sum_n \frac{\left(Re[\delta_h^{\star} \delta_m] -  \left(b_G + A_\mathrm{NG} \frac{2 f_\mathrm{NL}}{D(z_o) {\cal M}(k)}\right)\delta_m^{\star} \delta_m\right)^2}{{\cal P}_{mm}(k)/2\bar{n}} 
\end{equation}

Because there may be a slight dependence of $b_G$ on $f_\mathrm{NL}$, we consider two distinct fitting procedures for $A_\mathrm{NG}$.  In both, we assume the $k$ and $f_\mathrm{NL}$ dependence in Equation \ref{myNGbias} and define one and two sigma errors by changes in the $\chi^2$ value of 1 and 4 in Equation \ref{chi2eq}, where the sum is performed over the density modes of interest in each different $f_\mathrm{NL}$ simulation.  Following Ref. \cite{grossi/etal:2009}, we limit all fits to modes with $k \leq 0.03 h/$Mpc, which after discarding modes with wavenumber $2\pi/L_\mathrm{box}$ and $\sqrt{2} \times 2\pi/L_\mathrm{box}$, amounts to 366 modes.  We find that for the massive halos considered here, $\chi^2$ can be substantially smaller than the number of degrees of freedom (see \cite{smith/scoccimarro/sheth:2007,seljak/hamaus/desjacques:2009} for other recent evidence for sub-Poissonian sampling).  Therefore, our error bars may be overestimated.

In the first approach applied in the main text, we first fit the model with $A_\mathrm{NG}=0$ to the $f_\mathrm{NL} = 0$ simulation to determine $b_G$ and its uncertainty for each mass, redshift, and $z_f$-dependent halo subsample.  Next we fit for $A_\mathrm{NG}$ using the four $f_\mathrm{NL} \neq 0$ simulations, assuming that the scale-independent contribution, $b_G$, is independent of $f_\mathrm{NL}$.  Therefore we use the measurement in the $f_\mathrm{NL} = 0$ simulation to marginalize over a $b_G$ common to all values of $f_\mathrm{NL}$, integrating over the probability distribution $P(b_G)$ derived from the $f_\mathrm{NL} = 0$ result. 

\begin{figure}
  \centering
\includegraphics[scale=0.55]{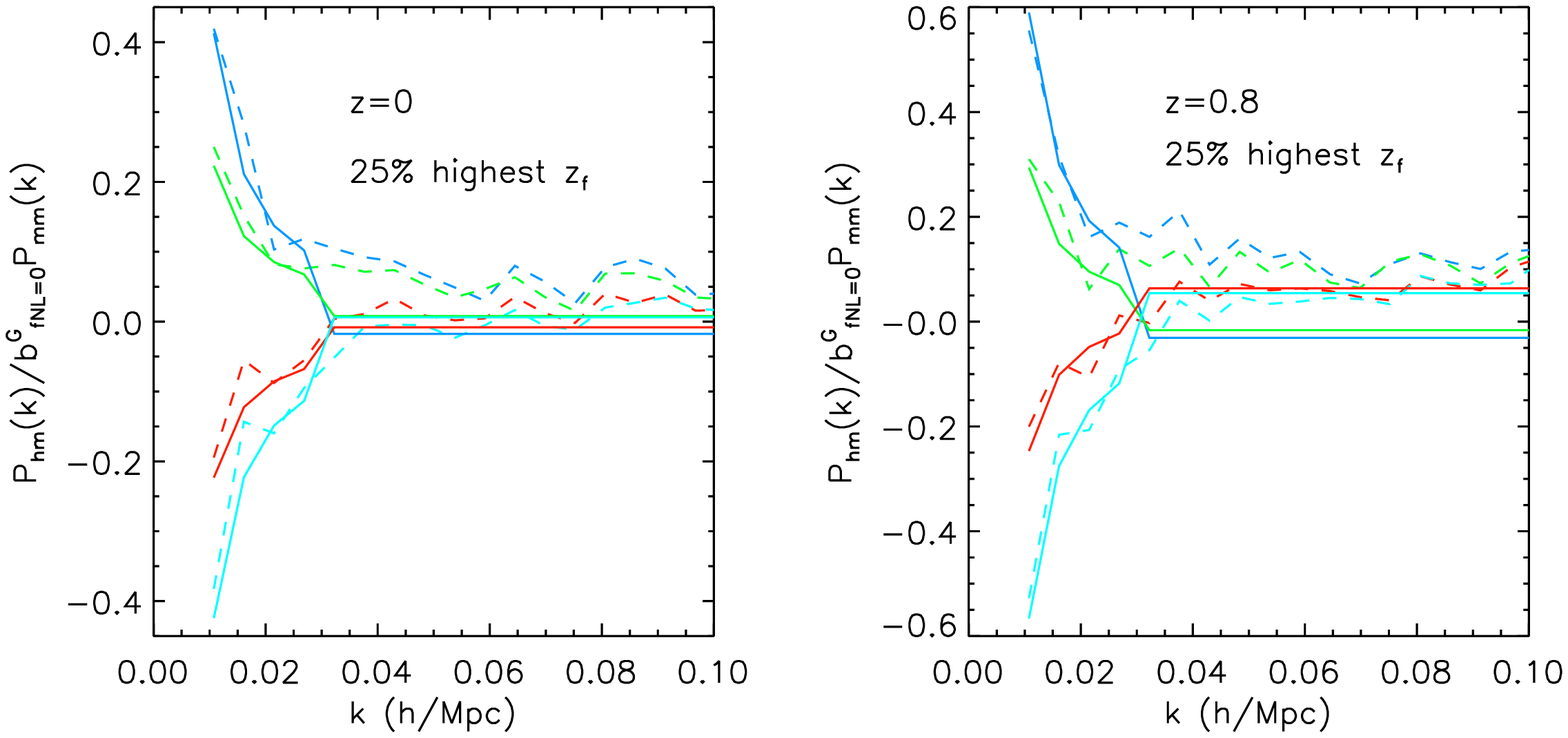}
  \caption{\label{fig:5pfit}  The colored dashed curves show $P_{hm}/b_{G, f_\mathrm{NL}=0} P_{mm} - 1$ for $f_\mathrm{NL}^\mathrm{LSS} = 200, 100, -100, -200$ (top to bottom, blue, green, red, light blue) for the high $z_f$ subsample with $M \geq 2\times 10^{13} h^{-1} M_{\odot}$ at $z=0$ (left panel) and $z=0.8$ (right panel).  The colored solid lines show the 5 parameter fit to the modes below $k=0.03 \; h$ Mpc$^{-1}$ for each value of $f_\mathrm{NL}^{LSS}$.  For $k > 0.03 \; h$ Mpc$^{-1}$ we show the best fit values of $b_G$ for each $f_\mathrm{NL}^{LSS}$ simulation.  In the right panel, $b_G$ appears anti-correlated with $f_\mathrm{NL}^{LSS}$.}
\end{figure}

\begin{figure}
\centering
\includegraphics[scale=0.55]{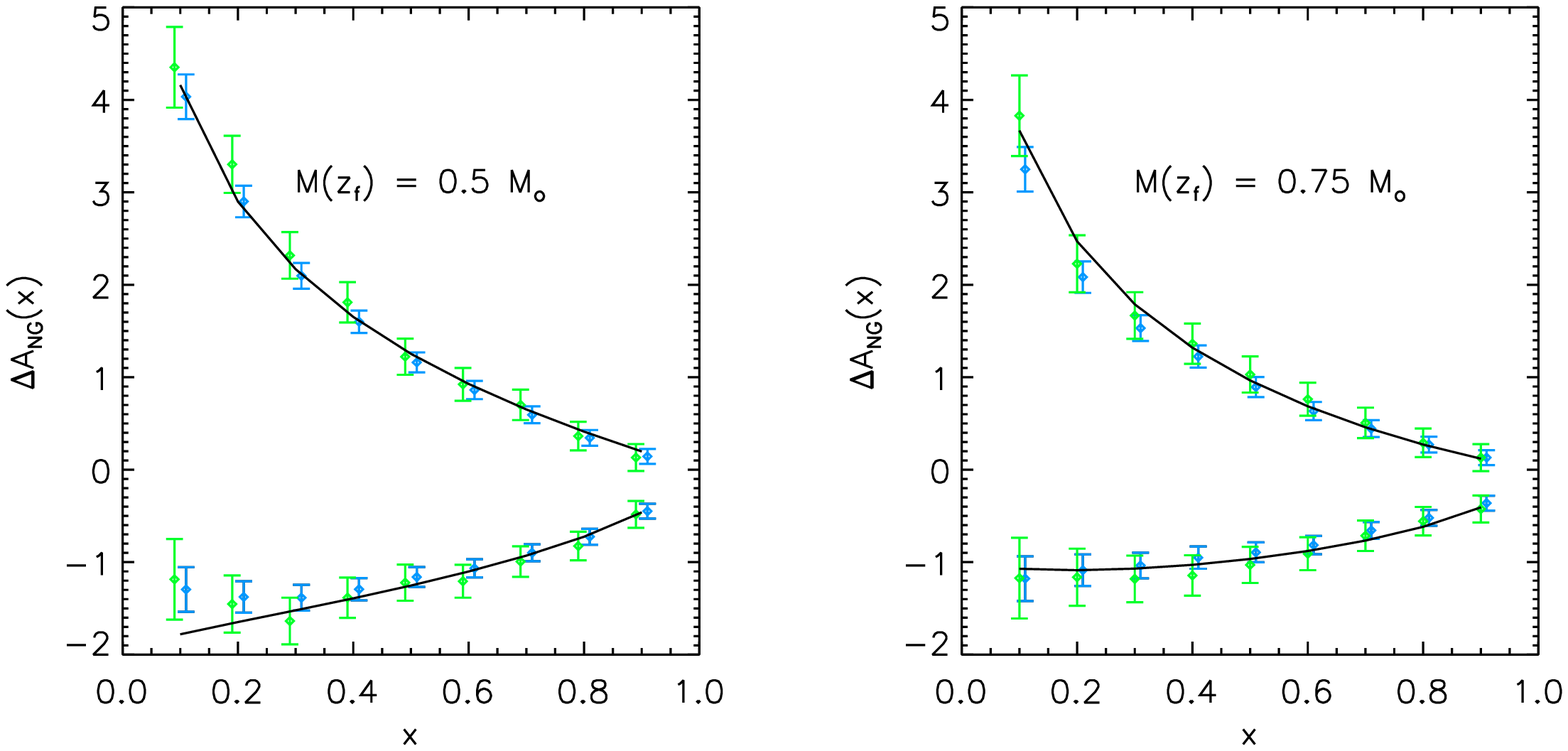}
\caption{\label{fig:5pvs1p}  Same as Figure \ref{fig:simsvsePS}, with $f=0.5$ in the left panel, and $f=0.75$ in the right panel.  We compare fits to $\Delta A_\mathrm{NG}$ with five parameters (green, larger errors) with one parameter (blue, smaller error bars and offset from $x$ by 0.01 for clarity) as described in the text.  The black curve shows the ePS prediction.}
\end{figure}

In the second, more conservative approach, we fit our simulation data to a five parameter model: the scale-independent $b_G$ in each non-zero $f_\mathrm{NL}$ simulation, and a single amplitude $A_\mathrm{NG}$ for the non-Gaussian signal.  Because of the limited range of $k$ values over which to fit, the non-Gaussian amplitude $A_\mathrm{NG}$ can be degenerate with the scale-independent bias $b_G$ in Equation \ref{deltahmodel}.  Figure \ref{fig:5pfit} shows an example of this.  We first fit the $f_\mathrm{NL} = 0$ simulation for $b_{G,f_\mathrm{NL}=0}$ in the same $k$ range we use to fit the non-Gaussian bias ($k \leq 0.03 h/$Mpc), holding $A_\mathrm{NG} = 0$.  To illustrate the non-Gaussian signal, we plot the halo-matter cross power spectrum ${\cal P}_{hm}$ normalized by $b_{G,f_\mathrm{NL}=0} {\cal P}_{mm}$ for $f_\mathrm{NL}^\mathrm{LSS} = 200, 100, -100, -200$ (blue, green, red, light blue).  We plot the best fit five parameter model in black for $k \leq 0.03 h/$Mpc, and we plot the value of $b_G$ for each $f_\mathrm{NL}$ value at $k \geq 0.03 h/$Mpc.  The best fit values of $b_G$ are anti-correlated with the value of $f_\mathrm{NL}$ in the $z=0.8$ sample, which could in principle affect the best fit value for $A_\mathrm{NG}$.  We compare the results of the five and one parameter fits to $A_\mathrm{NG}$ in Figure \ref{fig:5pvs1p}.  The two approaches give consistent results, though the errors are much larger as expected when $b_G$ is fit separately for each value of $f_\mathrm{NL}$.  

\begin{figure}
  \centering
\includegraphics[scale=0.55]{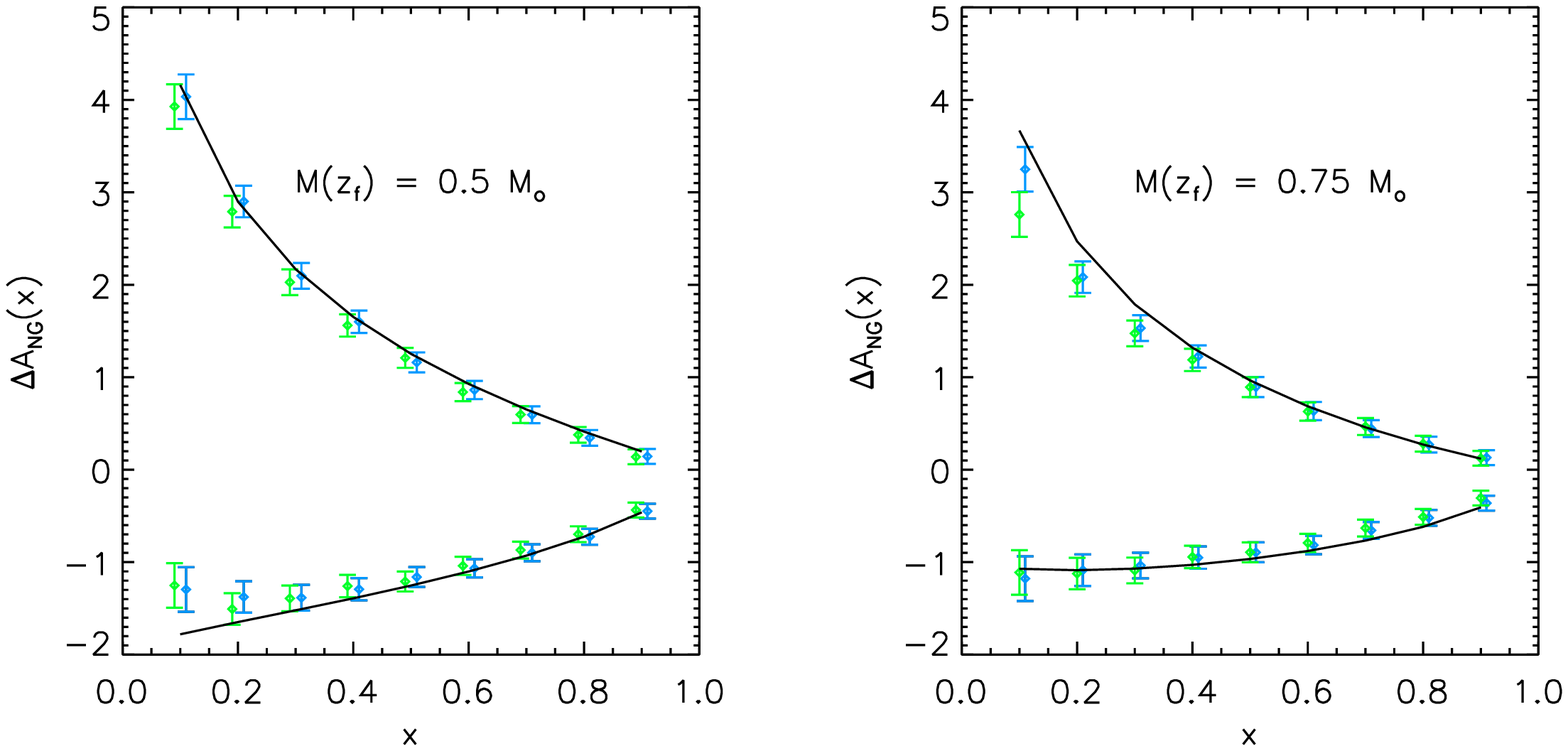}
  \caption{\label{fig:comparesplits}  $\Delta A_\mathrm{NG}(x)$ measured from subsamples defined with $N_\mathrm{group} = 100$ (green) and $N_{group} = 10000$ (blue).  The results are insensitive to this parameter entering how we select $z_f$-dependent halo subsamples with identical mass functions.  See the text for details.}
\end{figure}

In Section \ref{simresults} we introduce the parameter $N_\mathrm{group}$, which determines how many halos are grouped together before dividing into subsamples based on $z_f$.  If $N_\mathrm{group}$ is too small, one expects to introduce sample variance, in that halos are scattered across boundaries because each $N_\mathrm{group}$ set of halos is a finite sampling of the true $z_f$ distribution.  However, if $N_\mathrm{group}$ is too large, then one will introduce a spurious $z_f$ dependence through the dependence of $P_{z_f}$ on halo mass.  In Figure \ref{fig:comparesplits} we compare $\Delta A_\mathrm{NG}$ values for $N_\mathrm{group} = 100$ and $N_\mathrm{group} = 10000$, and demonstrate that our results are insensitive to this choice.
 
\end{document}